\documentstyle[11pt,newpasp,twoside,epsf]{article}
\begin{document}
\title{Magnetically powered prompt radiation and flow acceleration in GRB}
\author{H.C.\ Spruit  and G.\ Drenkhahn}
\affil{Max Planck Institute for astrophysics, box 1317, 85741 Garching, Germany}

\begin{abstract}
The physics of GRB powered by a magnetic energy flux is reviewed. Magnetic fields are natural for transmitting the energy from the central compact object to the small amount of baryons required for a GRB. When dissipation of the flux of magnetic energy by reconnection inside the flow is taken into account, the magnetic model assumes several more convincing properties. For baryon loading typical of observed GRB, most of the dissipation takes place just outside photosphere, so that prompt emission is produced efficiently, and the magnetic field strength in this region is high, resulting in efficient synchrotron emission. Remarkably, the dissipation also causes very efficient acceleration of the bulk flow. This effect is illustrated with a classical hydrodynamic equivalent. In this context, the distinction between the flux of magnetic energy $cB^2/8\pi$ and the Poynting flux $cB^2/4\pi$ is important, and an interpretation of the Poynting flux as a `magnetic enthalpy flux' illuminating. Numerical and analytical results for flow acceleration and the relative contribution of photospheric (thermal) and nonthermal emission as functions of the asymptotic bulk Lorentz factor are given. The transition between X-ray flashes and true GRB is predicted at $\Gamma\approx 100$.

\end{abstract}
 
\section{Introduction}

The prompt radiation of gamma-ray bursts accounts for a significant fraction of the total energy emitted. The original fireball model (Paczy\'nski 1986, Goodman 1986,  Meszaros \& Rees 1992) is successful in explaining the high Lorentz factor of the outflow, the short observed time scales, and the afterglow. For the amount of baryon loading inferred from time scales and total energy, however, the direct radiation from such a simple fireball is small, since almost all the energy is converted into bulk kinetic energy before it can be radiated at the photosphere. In addition, this radiation would be produced in an optically thick environment, hence quasi-thermal, in contrast with observations. 

The requirement to produce nonthermal $\gamma-$rays efficiently has given rise to the internal shock model (Rees \& Meszaros 1994), now a standard part of GRB interpretations. It takes into account that the light curves of GRBs are very erratic, indicating that the output from the central engine is also strongly variable. The relativistic outflow from such a modulated source is likely to have strong variations in flow speed. Shells of mass ejected at different speeds overtake each other, so that a fraction of the outflow energy dissipates in the form of internal shocks. To the extent that this dissipation takes place outside the photosphere it can plausibly produce nonthermal emission, seen as $\gamma$-rays by an external observer. 

The attraction of the internal shock model is the connection it makes between two key observations, the nonthermal prompt emission and the strong variability.  But the shells with the largest differences in speed (or Lorentz factor) overtake each other close to the central engine, in the optically thick part of flow. Their energy then thermalizes and adds to the density of the pair plasma, and through the expansion of the flow, this energy is converted mostly into kinetic energy. This increases the Lorentz factor, but does not add much to the observed nonthermal radiation. For modest modulations of the central engine (factors of a few) the fraction of the energy appearing as nonthermal radiation is only a few per cent (Daigne \& Mochkovitch 1998). The efficiency can be larger for large variations in bulk Lorentz factor (Beloborodov 2000). Such large variations in Lorentz factor, however, would also cause the spectrum of emitted radiation to vary dramatically during a burst, possibly much more than is observed (Fenimore, private communication). 

\subsection{Internal shocks vs. magnetic energy}

It may be hard to put firm observational limits on the efficiency at which the energy must be converted into prompt $\gamma$-rays, but values of a few tens of per cent would certainly be much easier to reconcile with observations than a few per cent. Such high efficiency would be most easily understood if the energy flux were in non-kinetic form and dissipated {\it outside} the photosphere, with only a fraction of the flux in kinetic energy, gives rise to the afterglow by interaction with the ISM or circumstellar medium. In the following we show that the Poynting flux produced by a nonaxisymmetric magnetic central engine has just the right properties (cf. also Thompson 1994).

\section{Magnetic central engines}
Powering a GRB requires the release of $10^{51}$ erg into a mass of $10^{-4}$ M$_\odot$.  This amount of energy and the observed short time scales require the source to be a compact object of a solar mass or more. Thus a mechanism is needed to transfer the energy released to a small amount of mass rather than spreading it out over the 1-10 M$_\odot$ contained in the source. Energy transport by neutrinos has been popular in connection with the merger scenario, but its efficiency in producing a fireball has so far been found to be too low (Ruffert and Janka, 1999). Strong magnetic fields, on the other hand, can couple a rapidly rotating object very effectively to a low density region surrounding it, through the `Poynting flux', carried by a magnetized outflow (for a review, see e.g. Blandford 2002). Rapidly rotating compact objects are the key in several types of central engine, including the neutron star + black hole model and some versions of the collapsar model (Woosley 1993, Paczy\'nski 1993), with precursor models in the form of the magnetically powered supernovae of LeBlanc and Wilson 1970, Bisnovatyi-Kogan 1976.

In the collapsar class of models, material with just the right amount of angular momentum forms a torus around the newly formed compact object in the core of a massive star. The differential rotation of this torus can generate a strong magnetic field, in the form of the magnetic turbulence known from numerical simulations of accretion disks (Hawley et al 1995). In a second class of models, the central object is a rapidly rotating neutron star. To make a GRB, this star has to be suddenly (seconds or less) magnetized to field strengths of the order  $10^{16}$ G. Proposed mechanisms for producing a sudden magnetization include accretion-induced collapse of a white dwarf (Usov 1992, Ruderman and Kluzniak 1998, Dai \& Lu 1998, Blackman \& Yi 1998) or the differential rotation produced by gravitational wave instability (Spruit 1999).

Common between these classes of models is a strong ($10^{15} - 10^{16}$) G magnetic field, 
rotating at subrelativistic speeds. Whether axisymmetric or nonaxisymmetric, such a rotating field produces a powerful magnetic energy flux.

\subsection{Magnetic energy flux and Poynting flux}

Though usually thought of as a property of vacuum waves, the Poynting flux $S=c{\bf E\times B}/4\pi$  is a more generally useful quantity. In ideal MHD\footnote{We assume ideal MHD here. It turns out that for typical GRB parameters (energy, mass flux) ample charge carriers are present to  maintain ideal MHD conditions ($E\ll B$ in the fluid frame) out to distances several orders of magnitude larger than the photosphere. One also verifies that the charge density which generically accompanies relativistic MHD flows only has fluctuating components, associated with the reconnection process. The charge density associated with the net azimuthal field that carries the magnetic energy flux vanishes.}, the electric field is ${\bf E=v\times B}/c$, where ${\bf v}$ is the fluid velocity. The Poynting flux is thus $ {\bf S=v_\perp} B^2/4\pi$, where $ {\bf v_\perp}$ is the velocity component perpendicular to the field lines. 

Thus the Poynting flux in ideal MHD can be seen as a magnetic energy carried by the flow. However, it is also clear that it is not just the magnetic energy density $B^2/8\pi$ that is being carried, since that accounts for only half of the Poynting flux. The other half can be accounted for by interpreting ${\bf S}$ as the flux of magnetic {\it enthalpy}, $w_{\rm m}=U_{\rm m}+P_{\rm m}$, where $U_{\rm m}$ and $P_{\rm m}$ are the magnetic energy density and the magnetic pressure, both equal to $B^2/8\pi$. This is similar to ordinary hydrodynamic flow, where the thermodynamic energy flux is given by the flux of enthalpy $w=U+P$, instead of just the flux of thermal energy $U$. This distinction between enthalpy- and energy fluxes plays a crucial role in the following, where we discuss the acceleration of the flow. 

\subsection{Dissipation of magnetic energy}

Through instabilities and reconnection, the magnetic energy in the flow can be converted into kinetic energy of the plasma or into fast particles, and from there into heat or radiation. The effects of this dissipation depend on how fast it takes place. If the dissipation is very fast so that it takes place close to the central engine, the magnetic energy flux is converted into a dense pair plasma, which then expands creating a classical fireball. The main use of the magnetic field in this case is to transfer rotational energy of the central engine to the outside, into the low-baryon environment needed for a gamma-ray burst. 

If the dissipation is sufficiently slow, on the other hand, most of the magnetic energy flux can dissipate {\it outside the photosphere} of the flow. The dissipation then takes place in an optically thin environment. Instead of thermalizing, the fast particle populations produced by the reconnection process then radiate their energy as synchrotron emission in the magnetic field of the outflow. 

At the same time, dissipation of magnetic energy reduces the total pressure and thus creates an pressure gradient that accelerates the flow outward. In this way, a magnetic central engine can provide {\it both} the acceleration of the flow {\it and} the dissipation outside the photosphere needed for efficient prompt radiation. Finally, under just the same conditions, the magnetic field outside the photosphere is large enough ($10^7$-$10^8$ G) for the most likely radiation mechanism, synchrotron emission, to be very efficient. Hence three crucial ingredients for prompt GRB emission are the natural result of magnetic dissipation  in a magnetically powered GRB outflow.

\section{Flow acceleration: centrifugal vs. dissipative}
\label{enth}
A rotating magnetic field can accelerate a flow by two distinct mechanisms. The most familiar of these is perhaps the magnetic slingshot process (Bisnovatyi-Kogan and Ruzmaikin 1976, Blandford \& Payne 1981, Lovelace 1976, for a tutorial introduction see Spruit 1996), which has been used successfully to explain outflows like the jets from protostellar objects and AGN. This process can take place without any thermal pressure or dissipation of magnetic energy. It can be visualized as a flow driven by the centrifugal force acting on matter tied to magnetic field lines which corotate with the central object. It works best in an axisymmetric configuration (relative to the rotation axis). With the magnetic field strengths and rotation rates needed to produce a gamma-ray burst, the centrifugal process can easily accelerate the flow to a Lorentz factor of the order 2.  Higher Lorentz factors are possible as well, but  are not achieved very efficiently: most of the energy flux tends to remain in magnetic form (e.g.\ Bogovalev 1997, Daigne and Drenkhahn 2001). This effect already seen in the first relativistic calculation by Michel (1969). 

The dissipation-induced flow acceleration mechanism (Drenkhahn 2002) does not depend on the centrifugal process, though centrifugal acceleration may also contribute to the early stages of acceleration of the flow, near the light cylinder. Instead, acceleration results from the {\it dissipation} of the magnetic energy in the flow. Energy dissipated by magnetic reconnection is converted into thermal energy (in the optically thick parts of the flow) or radiated away (in the optically thin parts). The simplest is the optically thin case. To illustrate it consider a plane parallel model, in which the flow $v(x)$ is in the $x-$ direction, and the magnetic field $B(x)$ perpendicular to it. 

Assume that the magnetic energy density $U_{\rm m}=B^2/8\pi$ declines on a length scale $L$, so that $dU_{\rm m}/dx=-U_{\rm m}/L$. The magnetic energy is converted into thermal energy, and from there into radiation. These steps are assumed to take place on a short time scale, so that the energy of the reconnecting field is lost quasi-instantaneously as nonthermal radiation.

Since the magnetic pressure is the same as the magnetic energy density, there is then a magnetic pressure gradient $dP_{\rm m}/dx=-U_{\rm m}/L$, which would be absent without dissipation. This gradient accelerates the mass in the flow. If the flow is already relativistic with Lorentz factor $\Gamma$, the acceleration is ${\rm d}\Gamma/{\rm d}t={\rm d}/{\rm d}t(U_{\rm m}/\rho_0 c^2)$, where $\rho_0$ is the rest mass density (as measured in the frame of the central engine). 

The acceleration is thus proportional to the rate of dissipation, and the kinetic energy attained by the flow equals the magnetic energy dissipated. It may seem strange that acceleration is possible by {\it destroying} the ingredient that drives it! To see how this can be, recall that the energy equation contains the Poynting flux $v_\perp B^2/4\pi$, i.e. twice the flux of magnetic energy. The effect of dissipation can be regarded as converting half of the Poynting flux into radiation or thermal energy, while the other half produces bulk acceleration of the flow.

\subsection{Hydrodynamic equivalent}
The somewhat anti-intuitive `acceleration by energy loss' process can also be clarified with the equivalent in ordinary, nonrelativistic hydrodynamics. Consider a one-dimensional, steady flow (in the $x$-direction) under the influence of a gas pressure only. If the flow is adiabatic (no dissipation or sinks of heat), the equation of motion $u\partial_x u=-\partial_x p$ can be integrated to yield the Bernoulli  equation, ${1\over 2}v^2+w={\rm cst} $,
where the enthalpy $w=u/\rho +p/\rho$, and $u$ is the thermal energy per unit volume of the gas. This is derived using the thermodynamic relation ${\rm d}w={\rm d}p/\rho \vert_S$, where the differential is taken at constant entropy $S$. The Bernoulli equation expresses energy conservation in the flow, such that the kinetic energy and a thermodynamic energy add up to a constant. 

In this one-dimensional, adiabatic, plane parallel case, the flow speed is constant. To obtain acceleration, one can appeal  to a diverging flow, as in the familiar example of a jet nozzle. A diverging flow is not the only possibility, however. Even in a plane parallel case the flow is also accelerated if it is not {\it adiabatic}. 

To see this, note that the thermodynamic energy involved is not the internal energy (thermal energy per unit mass) but the {\it enthalpy}, which has the additional contribution $p/\rho$. Where does this contribution come from? The Bernoulli equation was derived under the assumption of steady flow. The flow starts with finite speed and density at a source location $r_{\rm s}$ and flows out to infinity.  At $r_{\rm s}$ the energy is transferred from the source to the flow. This transfer includes two parts: the internal (thermal) energy {\it carried into} the flow region by advection, and the $P{\rm d}V$ work {\it done on} the flow at $r_{\rm s}$. The enthalpy thus comes in because of the nature of the interface between the source and flow regions. In a steady flow with energy loss due to radiation, only the internal energy is lost. There is still the $p{\rm d}V$ work done by the source, and this is what accelerates the flow.

For a bubble of energy shot out instantaneously, the contact between source and flow does not last, so the situation is then somewhat different. For typical GRB parameters, however, the flow time from the central engine to the photosphere (as seen by the observer) is of the order of a millisecond, so that the steady approximation is relevant for all but the very shortest time scales seen in bursts. 

\section{Dissipation by reconnection} 
Due to the rotation of the source, the magnetic field in the outflow is highly `wound up' into an almost purely azimuthal field. In the ultrarelativistic limit, this field just flows out radially at the speed of light, so that the field strength varies as $1/r$. To next order in $1/\Gamma^2$, the field can deviate from this, because it can evolve internally in the flow. Such evolution can take place by instabilities and by reconnection. If the source of the field is mostly axisymmetric, kink instabilities (Eichler 1993, Spruit et al. 1997) can reduce the field energy and convert it into kinetic energy of internal motion, and ultimately into heat and radiation. If the source is mostly non-axisymmetric, the outflow is  a `striped wind' (Kennel and Coroniti 1984, Bogovalov 1999): the field lines change direction with distance from the source on a length scale (as seen in a the frame of the central engine) $\lambda=2\pi c/\Omega$, i.e. of the order of the light cylinder radius (100km for typical central engine parameters). This length is quite short compared with, say, the photosphere radius of a GRB. 

Nearby field lines of different direction tend to reconnect, decreasing the magnetic energy by converting it into flows (MHD waves) and fast or thermal particles accelerated by electric fields in the reconnection region. Is this likely to be relevant in magnetic GRB outflows? To estimate this, assume that neighboring field lines reconnect into each other at a speed $v_{\rm r}=\epsilon v_{\rm A}$, where $v_{\rm A}$ is the Alfv\'en speed. For the moment this scaling is only a formal parametrization, but below we will argue that $\epsilon$ is in fact a number of order unity. The time scale for reconnection is then, in a frame comoving with the flow
\begin{equation} 
\tau_{\rm r}^\prime=\lambda^\prime/v_{\rm r}^\prime=\Gamma 2\pi P{c\over \epsilon v_{\rm A}^\prime},
\end{equation}
where $P$ is the rotation period of the source and we have used the Lorentz transformation $\lambda=\lambda^\prime/\Gamma$. Transforming $\tau_{\rm r}^\prime$ to the frame at rest with respect to the central engine introduces another factor $\Gamma$ due to time dilatation, $\tau_{\rm r}=\Gamma\tau_{\rm r}^\prime$. As long as the field has not decayed by reconnection, the Alfv\'en speed in the comoving frame is close to the speed of light, since the Poynting flux is so high compared with the flux of rest mass energy in a magnetically driven GRB. One thus finds
\begin{equation}  \tau_{\rm r}\approx\Gamma^2 P/\epsilon\end{equation}
i.e. the decay time of the field energy by reconnection scales with the rotation period of the central object. The rotation period is quite short, but the decay, as seen in the lab frame, is slowed down by the huge factor $\Gamma^2\sim 10^4-10^5$ due to the relativistic nature of the flow. 

\subsection{Reconnection rate}
Magnetic reconnection has been studied especially intensively in special field configurations such as two-dimensional, initially exactly antiparallel fields. The reconnection rate in such well-ordered fields is known to depend on details such as boundary conditions applied. In a magnetic outflow such a well-organized field configuration develops naturally if no reconnection were to take place: the field quickly becomes almost azimuthal, with regular changes of direction on the length scale $\lambda$. This conceptually simple picture has led to the development of `striped wind' reconnection models for pulsars (Kennel and Coroniti 1984). Near the light cylinder, however, the field is less simple because the azimuthal component does not yet dominate, and any small scale structure of the field at the source still noticeable. This field is likely to be dynamically evolving from the beginning. A smooth `striped wind' may in fact never develop. 

We take the opposite point of view here: assume that the field in the outflow is in fact in a dynamic state of instability and reconnection from the beginning. The evolution time scale is then the Alfv\'en time scale across the natural length scale $\lambda$. As in the case of hydrodynamic flows, and in the magnetic turbulence observed in simulations of accretion disks (Hawley et al. 1995), we guess that the reconnection rate is effectively determined not by the microscopic diffusion processes but by the dynamical time scale. As long as diffusion is unimportant, the large scale dynamics creates small scales on a dynamical time scale, until the reconnection rate matches overturning rate on the small scale. The evolution time scale of the field would then be independent of the microscopic processes (perhaps up to a factor of order $\ln R_{\rm m}$. This would be similar to high-Reynolds number flows in hydrodynamics, where the (statistical) dynamics on large scales is often argued to be independent of the viscosity.

Cases of such a `disordered' form of magnetic reconnection are known. It is indicated indirectly by observations of the solar corona for example, and in this form has been reproduced in numerical experiments (Gudiksen et al. 2002). 

In this picture, the reconnection time scale is scales directly with the Alfv\'en time scale and we may guess a coefficient $\epsilon$ of order unity, perhaps 0.1.
In any case, irrespective of the above picture, the reconnection speed must generically approach the speed of light, in the limit of a highly relativistic magnetic field (magnetic energy density much larger than the rest mass energy density). This has been verified for a specific reconnection model by Lyutikov and Uzdensky (2002). The conditions in a GRB outflow, if it is driven magnetically, are well within this limiting case. Fortunately, the rate of reconnection is therefore much {\it less} uncertain than in the nonrelativistic case.
 
\subsection{Prompt radiation by magnetic dissipation}
The distance $r_{\rm diss}$ at which the decay of field energy becomes effective is $r_{\rm d}=c\tau_{\rm r}$. If this is smaller than the photosphere radius, $r_{\rm phot}$, the magnetic energy flux decays in the optically thick part of the flow. We call this the optically thick case. The energy dissipated then thermalizes, and on passing through the photosphere an approximately thermal spectrum is produced\footnote{Since electron scattering is likely to be important near the photosphere, thermalization actually takes place deeper inside the flow. The emergent spectrum depends on details of the dissipation and radiation processes between the thermalization depth and the photosphere.  In addition, photon upscattering by random motions associated with reconnection in this region (Thompson 1994) may be important. Thus it is quite possible that the spectrum produced deviates substantially from a black body even  in the optically thick case.}. The amount of radiation produced decreases with $r_{\rm diss}/r_{\rm phot}$: as in a basic fireball model, the radiation energy decreases rapidly by adiabatic expansion after dissipation has ceased. It is converted into kinetic energy of expansion, to reappear as radiation only in the afterglow. In the limit of very rapid dissipation of the magnetic energy, the magnetic dissipation model thus reverts to the basic hydrodynamic fireball model of Paczy\'nski (1986) and Goodman (1986).

The optically thin case obtains when most of the magnetic energy is dissipated outside $r_{\rm p}$. The radiation produced then depends on how the dissipation takes place. Fast electron distributions are expected to result from magnetic reconnection, but the bulk of the dissipated energy might simply appear in thermal electrons, such as is observed for example in solar flares. Since the magnetic energy has not yet disappeared but is in fact in the process of dissipating, the field strength is still high. For typical GRB parameters, the field strength at the photosphere will be of the order  $10^7-10^8$ G, so that synchrotron radiation is produced very effectively in the soft X-ray range even by mildly relativistic particles (and Doppler-boosted to the BATSE range by the outflow). Before a comparison with observed burst spectra can be made, however, a model needs to be developed for the particle distributions resulting from reconnection under the present relativistic conditions.

\section{Results}
\subsection{Flow speed}
With magnetic dissipation scaling with the Alfv\'en speed (and hence a fraction of the speed of light) as discussed above, the acceleration of the flow can be computed in detail, including the gas and radiation pressures where relevant. The results of detailed numerical calculations were reported in Drenkhahn and Spruit (2002). Starting near the light cylinder $r_{\rm l}$, the flow speed as a function of distance converges rapidly to a profile that depends only on the mass loading and energy flux. Acceleration stops at some large distance $r_{\rm f}$ when (almost all) the magnetic energy has dissipated. In the intermediate range, the equations simplify and an analytic solution is possible (Drenkhahn 2002). The flow in this intermediate range is approximately selfsimilar, and the Lorentz factor depends on distance as
\begin{equation} 
\Gamma=\left ( {3\over \pi c}\epsilon\Omega\Gamma_\infty r\right )^{1/3}, \qquad (r\gg r_{\rm l},~\Gamma \ll \Gamma_\infty) 
\end{equation}
where $\Gamma_\infty$ is the asymptotic Lorentz factor at large distance.
This dependence on distance differs somewhat from that in the model by Lyubarski and Kirk (2001) for the Crab pulsar wind, on account of a different prescription used for the reconnection speed. 

A remarkable detail is that flow acceleration by magnetic dissipation is almost as efficient in the optically thick part of the flow as it is in the optically thin region. In the optically thick regime, the dissipated magnetic energy stays in the plasma, so one might think magneti c dissipation would not create much of a pressure gradient. Actually, the conversion of magnetic energy into radiation is accompanied by a large drop in pressure. This is because the ratio of pressure to energy density in the magnetic field is unity, while in the radiation component it is $P_{\rm rad}/U_{\rm rad}=1/3$. The acceleration of the flow in the optically thick, radiation dominated regime is therefore only 33\% less efficient than in the optically thin case. To a first approximation, the difference between the optically thick and thin regimes can therefore be ignored when computing the acceleration resulting from magnetic dissipation.

\subsection{Conversion efficiency}
The initial Poynting flux is converted partly into radiation, the remainder into bulk kinetic energy, while a small amount remains in the flow as magnetic energy. In the optically thick case ($r_{\rm diss}\ll r_{\rm phot}$) almost all energy is converted to kinetic energy of the flow, as in the basic fireball model. In the optically thin case ($r_{\rm diss}\gg r_{\rm phot}$) a fraction is converted into prompt radiation and the rest into kinetic energy. The numerical results show that (for a purely nonaxisymmetric magnetic field at the source) the two contributions are equal, i.e. 50\% of the total energy flux is radiated as prompt emission. This can be understood in terms of the discussion above, by looking at the Poynting flux as a flux of magnetic enthalpy. Half of the enthalpy is given by the internal energy, which is radiated away in the optically thin case, while the other half accelerates the flow. 

\subsection{Thick-thin transition: X-ray flashes}
The results show (Fig 1) that the transition between optically thick and optically thin magnetic dissipation occurs when the terminal Lorentz factor is of order  $\Gamma_\infty = \Gamma_{\rm c}\approx 100$. For lower baryon loading, the Lorentz factor is larger and the dissipation takes place outside the photosphere. From the expressions for the photosphere radius $r_{\rm phot}=\dot E\kappa/(8\pi\Gamma^3 c^3)$ and the dissipation radius $r_{\rm diss}$, an approximate value for $\Gamma_{\rm c}$ is found (assuming electron scattering for the opacity $\kappa$):
\begin{equation}  
\Gamma_{\rm c}=100\, (\epsilon_{-1}\Omega_4\dot E_{50})^{1/5}.\label{gac}
\end{equation}
whereas the photosphere radius is
\begin{equation}  
r_{\rm phot}=10^{11}\,{\rm cm}\, (\epsilon_{-1}\Omega_4)^{-2/5}\dot E_{50}^{3/5}.
\end{equation}
 
For the bulk Lorentz factors inferred from observed GRB, $\Gamma_\infty \ga 100$. This is quite consistent with the magnetic dissipation model: for these values the dissipation takes place mostly outside the photosphere. Note also (eq. 4) that this value is rather insensitive (1/5th power) to the main uncertainty in the  magnetic dissipation model, the dimensionless reconnection rate $\epsilon$. A quasithermal component from the photosphere becomes dominant at $\Gamma_\infty \approx 100$, while at even higher baryon loading $\Gamma_\infty < 100$. This is the range where the model predicts the transition from GRB first to X-ray rich GRB and then to X-ray flashes to take place (Fig. 1). The X-ray flash range is limited: at even higher baryon loading, the efficiency of both thermal and nonthermal emission decreases quickly since all the energy is released inside the photosphere, and by expansion is converted into kinetic energy of the flow. As in the simplest hydrodynamic fireball model, such cases would yield `orphan afterglows'.

\begin{figure}
\mbox{}\hfill\epsfxsize 0.6\hsize\epsfbox{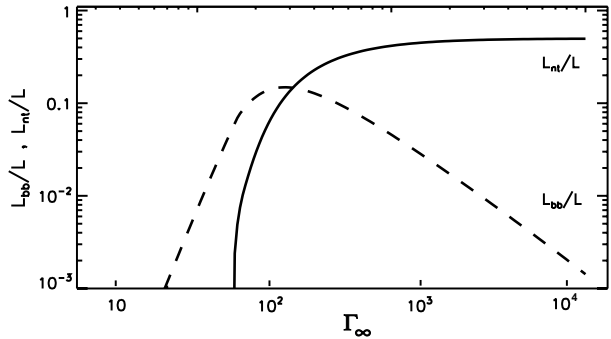}\hfill\mbox{}
\caption{
Thermal (photospheric, dashed) and nonthermal (solid) contributions to the prompt GRB emission predicted from the magnetic dissipation model, as a function of the asymptotic Lorentz factor of the flow. At large baryon loading (small $\Gamma_\infty$) prompt emission is inefficient and most of the total energy released is converted into kinetic energy. For $\Gamma_\infty$ around 100, X-ray flashes or X-ray-rich GRB result. For $\Gamma_\infty\gg 100$, 50\% of the total energy is emitted as prompt emission and 50\% ends up as kinetic energy (from Drenkhahn and Spruit, 2002).} 
\end{figure}

\subsection{Collimation?}
The magnetic dissipation model as presented here does not address the question of collimation of GRB outflows. The opening angles inferred from observations, of the order $10^{-2}-10^{-3}$ sr, are a key constraint on any model. Self-collimation is a popular idea in magneto-centrifugal models of jet acceleration, but it works less well in flows that are relativistic already close to to the generating object. If internal MHD instability in a self-collimated outflow is taken into account, high degrees of collimation become problematic even for non-relativistic flows, as we have argued elsewhere (Spruit et al. 1997). Though a possible way around this problem for highly relativistic outflows has been proposed by  Lyutikov and Blandford (2002), we conservatively view the processes described here as leading at best to only mild self-collimation. Collimation must then be done by something else, and the most natural candidate is the mechanism that is part of the standard `collapsar' model for the GRB progenitor (Woosley 1993). In this picture, a central magnetic rotating torus produces a weakly collimated energy flux, which is then collimated by the structure of the rotating progenitor envelope.
 
\section{Summary: what magnetic fields can do for GRB} 
One of the original motivations for exploring magnetic central engines for GRB was the ability of magnetic fields to carry energy across vacuum: they can in principle produce GRB with extremely low amounts of baryon contamination. This is a major advantage, but as we have shown here, the magnetic dissipation model has three more convincing properties:
\begin{itemize}
\item{-} The region where the magnetic energy flux dissipates is naturally located just outside the photosphere, causing it to appear as prompt nonthermal emission at an efficiency of 50\% of the total GRB energy,
\item{-} The dissipation process also accelerates the flow to the observed Lorentz factors,
\item{-} The high magnetic field strength in the dissipation region leads to efficient synchrotron emission,
\end{itemize}
while in addition the transition between the optically thick  (X-ray flash, high baryon loading) and optically thin emission (GRB, low baryon loading) takes place at an observationally plausible Lorentz factor around 100. 

This shows that dissipation of the magnetic energy flux from a nonaxisymmetric rotator  provides, in a natural way, the key elements of a successful model for GRB outflows. 
Whether on account of these intrinsic strengths, or by elimination of other possibilities, magnetic powering of GRBs is likely to become an important development in GRB modeling.

\end{document}